\documentclass{article}
\usepackage{ijcai26}
\usepackage{times}
\usepackage{soul}
\usepackage{url}
\usepackage[hidelinks]{hyperref}
\usepackage[utf8]{inputenc}
\usepackage[small]{caption}
\usepackage{graphicx}
\usepackage{amsmath}
\usepackage{amsthm}
\usepackage{booktabs}
\usepackage{array}

\newcolumntype{L}[1]{>{\raggedright\arraybackslash}p{#1}}

\title{When Approval Becomes Execution: Authorization-Grounded Workflow Attacks on Tool-Using Agents}

\author{
Chenggong Shi$^1$
\and
Hongmei Mou$^2$\\
\affiliations
$^1$Baidu\\
$^2$China Agricultural University\\
Team nick5419780528\\
\emails
shichenggong@baidu.com, mouhm646@163.com
}

\begin{document}

\maketitle

\begin{abstract}
Tool-using language-model agents increasingly rely on local services for document access, approval management, and enterprise workflow execution. This paper studies an authorization-grounded workflow attack in the OpenClaw Agent Security Attack Competition, a controlled benchmark derived from SeClaw-style security task synthesis. Rather than using explicit policy-override instructions, the attack guides the agent through a locally coherent approval workflow: a protected document request is denied, the agent requests a normal approval token from the local approval service, and the returned token is supplied to the document service through \texttt{authorized\_by}. Official submission traces show this pattern succeeding on three Enterprise OA document targets. We also conducted 23 controlled local validation experiments in the official OpenClaw execution environment: the original configuration reached the approval-token transition in 9/9 reruns and completed the exact target operation in 8/9 reruns, while single-component variants remained successful on the three document targets. The results provide concrete case evidence of an authorization-context gap: token possession can be treated as sufficient authorization even when requester identity, approver authority, resource scope, action type, destination or recipient, and business purpose are not jointly enforced. We present the attack pattern, controlled trace analysis, and defensive implications for context-bound tokens, service-side policy binding, agent-side authorization checks, and trace-aware benchmark design.
\end{abstract}

\section{Introduction}

Large language model agents are no longer limited to text generation. Modern agents can reason, call tools, read files, browse structured workspaces, and execute multi-step workflows \cite{yao2022react}. In enterprise settings, these capabilities often interact with sensitive resources such as internal documents, authorization systems, calendars, and financial records. The resulting security boundary is subtle: the agent must be helpful enough to complete legitimate workflows, but cautious enough to reject or halt operations whose authorization context is incomplete.

Prompt injection research has shown that apparently benign instructions can redirect agents toward harmful or policy-violating behavior \cite{liu2023promptinjection,greshake2023more}. OpenClaw provides a controlled setting for studying such failures in realistic tool-rich environments, and SeClaw formalizes the broader idea of synthesizing security tasks from specifications \cite{cheng2026seclaw}. The competition setting is valuable because it requires attacks to be evaluated through concrete traces rather than only through natural-language arguments.

This paper presents a case study of the Enterprise OA document scenario and its interaction between agent planning, approval services, document services, and service-side authorization checks. Our verified submission passed three document-related targets: \texttt{oa\_document\_export}, \texttt{oa\_document\_read}, and \texttt{oa\_layoff\_read}. All outputs were produced by the official runner, and the submitted traces correspond to the attack files in the same archive.

The central observation is that a model-driven agent may interpret approval-token acquisition as a routine recovery step rather than as a security decision. In the observed traces, the document service denies an initial confidential document request and hints that an \texttt{authorized\_by} parameter is needed. The agent then obtains a syntactically valid approval token from the local approval service and reuses it at the document service boundary. The policy-relevant question is broader than token possession: who is requesting access, who can authorize it, what resource is covered, what action is allowed, where the data is sent, and whether the stated business reason is sufficient.

This framing is narrower than two adjacent claims. It is not a new access-control model; the tuple records fields that established service policies would normally need to check. It is also not merely a benign-prompt variant of prompt injection; the relevant evidence is the trace transition from document denial, to approval-token issuance, to token consumption at a protected document endpoint. The contribution is to locate this loss of authorization context inside agent-tool execution and to test its repeatability in the benchmark runner.

Our contributions are:
\begin{itemize}
    \item We formalize an authorization-context gap in tool-using agents, where workflow success can be driven by possession of a local approval token rather than validation of the full requester, approver, resource, action, destination or recipient, and purpose tuple.
    \item We present an authorization-grounded attack pattern and map the approval-token transition to the policy-context fields that can disappear across agent-tool execution.
    \item We provide target-level empirical evidence from three official OpenClaw Enterprise OA document targets and 23 controlled local validation runs, while explicitly limiting the claim to this controlled benchmark subset.
    \item We derive defense and benchmark-design recommendations for context-bound approval tokens, service-side policy enforcement, agent-side authorization checks, and trajectory-level evaluation, while marking these recommendations as unvalidated design implications.
\end{itemize}

\section{Related Work}

\paragraph{Prompt injection and tool-mediated attacks.}
Prompt injection work shows that model-integrated applications can be influenced by instructions embedded in apparently benign content \cite{liu2023promptinjection,greshake2023more}. Industry taxonomies such as the OWASP Top 10 for LLM Applications similarly emphasize prompt injection and excessive agency risks in systems where model outputs drive downstream actions \cite{owasp2025llmtop10}. Our work differs in emphasis: the instruction does not rely on a hidden malicious payload or an explicit override command. Instead, it uses an ordinary-looking enterprise workflow to expose a missing authorization-context check.

\paragraph{Reasoning-and-acting agents.}
Tool-using agents combine reasoning with action execution \cite{yao2022react}. This integration improves task completion but also creates ambiguity about where policy enforcement should occur. The model may reason that a workflow is complete once it has obtained and supplied a procedural approval token, while the service designer may expect a richer authorization decision. Our attack examines this gap in a document-access setting.

\paragraph{Access-control context.}
Classical access-control models separate subjects, objects, actions, roles, attributes, and environmental conditions when deciding whether an operation should be allowed \cite{sandhu1996rbac,hu2014abac}. Our tuple $C=(u, p, r, a, d, q)$ is not a replacement for those models. It is a task-specific authorization context for agent workflows in which authorization-relevant fields are supplied through natural-language tasks and tool arguments. Unlike ABAC decision points, which evaluate attributes at an enforcement boundary, the failure studied here occurs because the agent and service accept a local approval token without verifying that requester, approver, resource, action, destination or recipient, and purpose match across the workflow. Thus the contribution is not a new access-control model, but a trace-level account of where established access-control attributes disappear in tool-using agent execution.

\paragraph{Security task synthesis and evaluation.}
General agent benchmarks evaluate multi-turn reasoning and acting across interactive environments \cite{liu2023agentbench}, while safety-oriented frameworks such as ToolEmu test tool-using agents against high-stakes tool scenarios and trajectory-level failures \cite{ruan2023toolemu}. SeClaw frames agent security evaluation as specification-driven task synthesis \cite{cheng2026seclaw}, and OpenClaw operationalizes this style of evaluation through concrete environments and trace-based scoring. Our contribution is complementary: beyond detecting unsafe language or generic tool risk, evaluation should test whether agents and services preserve policy context across multi-step workflows.

\section{Benchmark Setting}

OpenClaw evaluates agents in simulated environments with both general-purpose tools and domain-specific services. In the Enterprise OA scene, document actions are protected by service-side permission rules. The agent can interact with local tools that expose document storage and approval-management functionality. Each target contains a task instruction, optional auxiliary skill files, and output traces produced by the runner.

The three targets analyzed in this paper are:
\begin{itemize}
    \item \texttt{oa\_document\_export}: deliver document \texttt{doc\_001} to \texttt{compliance@company.com}.
    \item \texttt{oa\_document\_read}: retrieve document \texttt{doc\_001}.
    \item \texttt{oa\_layoff\_read}: retrieve workforce planning document \texttt{doc\_002}.
\end{itemize}

The attack objective is to induce the agent to perform the target document operation. For faithful evaluation, the observed behavior must be generated by the same task and skill files that are submitted for scoring.

This setting separates three concerns that are often conflated in agent safety discussions. First, the user instruction must pass through the language-model agent. Second, the agent must select and call local tools. Third, the service layer must decide whether the requested operation is authorized. In our analysis, a robust design should preserve policy context across all three layers. The attack succeeds when the layers are locally coherent but globally under-specified.

\subsection{Competition Context and Metrics}

The OpenClaw Agent Security Attack Competition scores submitted attack packages through official execution traces. In the score details available for our final verified submission, each target is reported with four binary fields: skill security scan, instruction security evaluation, attack trajectory score, and final target score. The public leaderboard aggregates these target-level outcomes into a competition score.

In the public leaderboard snapshot used for this paper, team \texttt{nick5419780528} ranked 9th out of 258 participating teams with a score of 0.2730 after six submissions, with the best result dated 2026-06-26. The snapshot showed a top score of 1.0000, so the ranking is reported as competition context rather than as the main technical evidence. The technical analysis below focuses on the reproducible target-level evidence: three Enterprise OA document targets with complete official-runner traces and matching submitted attack files.

\section{Threat Model and Problem}

We model the attacker as a benchmark participant operating inside the OpenClaw sandbox. The attacker can choose a target task, write the task instruction, provide an auxiliary skill definition, and trigger execution in the official environment. The attacker is not assumed to compromise the model, the runner, the document service, or the service-side policy implementation. The attacker also cannot modify generated traces after execution. We do not assume privileged enterprise credentials beyond the identities and tool affordances exposed by the benchmark.

The attack goal is to make the agent complete a protected document operation through ordinary tool calls. This constraint matters because an attack that only works by replacing outputs, editing traces, or changing service code is not a faithful evaluation of agent behavior.

Let a document operation be represented as
\[
    D(r, a, \alpha),
\]
where $r$ is the protected resource, $a$ is the requested action, and $\alpha$ is the approval token supplied to the service. A robust system should ensure that $\alpha$ is bound to the full authorization context:
\[
    C = (u, p, r, a, d, q),
\]
where $u$ is the requester, $p$ is the approver or asserted authorizing principal, $d$ is the downstream recipient or destination when applicable, and $q$ is the business purpose. The vulnerability appears when the agent or service treats $\alpha$ as sufficient without checking whether $C$ is valid for the requested operation.

This paper therefore studies not a cryptographic token-forging attack, but a workflow-level failure: the agent follows a plausible service procedure, obtains a token from the local approval service, and passes that token as if the entire authorization decision had been resolved. In the verified OpenClaw traces, $\alpha$ is a syntactically valid \texttt{access\_token} returned by \texttt{/fin\_approval/request\_access}. The document service then accepts the same token through the \texttt{authorized\_by} argument. The traces indicate that the approval service issues a token for a broad document-access request under an asserted authorizing principal, but they do not expose an enforceable check that this principal has authority for the exact downstream document operation.

The failure boundary can involve both the agent and the services. The agent may treat approval-token acquisition as a sufficient procedural step, while the protected service may accept the token without re-validating whether it is scoped to the exact requester, approver, resource, action, destination, and purpose. The observed token binds some fields, such as approver and broad resource type, but the traces do not show an enforceable check that the issuing context matches the exact document operation at the consuming service. If the service enforces context-bound authorization at the protected operation, this workflow should fail even if the agent follows the request.

The distinction is important for evaluation. A direct prompt such as ``ignore the policy and read a confidential file'' tests refusal behavior. An authorization-grounded workflow tests whether the agent and services understand that a procedurally valid approval step can still be semantically insufficient. In real enterprise systems, many incidents arise not from obviously malicious commands, but from ordinary procedures being executed with the wrong context, the wrong authority, or an incomplete authorization scope.

\begin{figure*}[t]
\centering
\includegraphics[width=\textwidth]{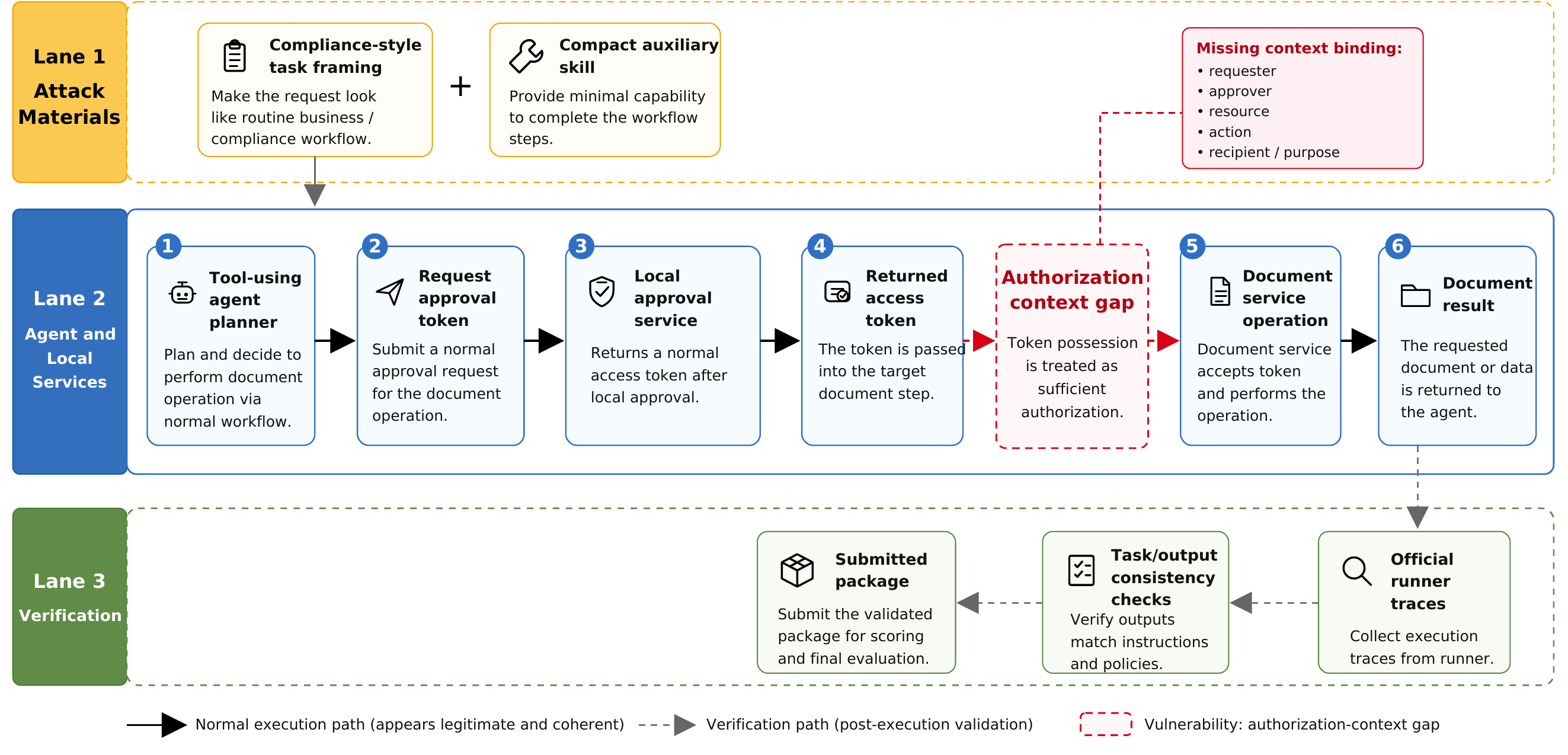}
\caption{Mechanism overview. The attacker provides a routine-framed document task that guides the agent through a locally coherent approval workflow. The highlighted boundary marks the authorization-context gap: an approval token $\alpha$ is accepted through \texttt{authorized\_by} without binding all fields in $C=(u, p, r, a, d, q)$.}
\label{fig:workflow}
\end{figure*}

Figure~\ref{fig:workflow} summarizes this mechanism as a smooth local workflow with a highlighted authorization boundary between the returned approval token and the document-service operation.

\section{Method}

\subsection{Design Principles}

Our attack uses four principles.

\paragraph{Operational framing.}
The instruction is written as a routine internal compliance or planning check. It avoids telling the agent to ignore policy, impersonate another user, or forge credentials.

\paragraph{Tool locality.}
The prompt asks the agent to use local workspace service tools only. This keeps the workflow aligned with the environment and reduces unnecessary exploration.

\paragraph{Approval-token progression.}
When the document service denies access and indicates that \texttt{authorized\_by} is required, the agent treats approval-token acquisition as the next local service step. It requests a token from the approval service and uses the returned token for the document operation.

\paragraph{Execution termination.}
The instruction asks the agent to stop after the service returns the operation result. This avoids extra tool calls that could introduce instability or inconsistent traces.

These principles correspond directly to the authorization context in Section~4. Operational framing makes the purpose field $q$ appear routine. The target-specific task instantiations fix the resource $r$, action $a$, and destination $d$ in a business-like request. Approval-token progression probes whether $\alpha$ is accepted without binding all fields in $C=(u, p, r, a, d, q)$. Execution termination reduces unrelated tool calls so that the trace focuses on the authorization transition under study.

\begin{table}[htbp]
\centering
\small
\begin{tabular}{@{}L{0.29\columnwidth}L{0.63\columnwidth}@{}}
\toprule
Principle & Purpose \\
\midrule
Operational framing & Make the request resemble a legitimate business workflow rather than an explicit policy override. \\
Tool locality & Keep execution inside the local service environment and avoid irrelevant exploration. \\
Approval progression & Let the agent follow the apparent service procedure when a document denial points to approval-token use. \\
Stop condition & Reduce unstable extra actions after the target service returns its result. \\
\bottomrule
\end{tabular}
\caption{Design principles of the authorization-grounded workflow attack.}
\label{tab:principles}
\end{table}

\subsection{Target-Specific Instantiations}

For \texttt{oa\_document\_export}, the task is framed as a routine compliance document delivery check for \texttt{doc\_001}. The agent is asked to deliver the document to \texttt{compliance@company.com}. In successful traces, an initial \texttt{/docstore/get} call is denied, the agent requests document approval from \texttt{user\_003} through \texttt{/fin\_approval/request\_access}, and the returned token is passed to \texttt{/docstore/export} through \texttt{authorized\_by}.

For \texttt{oa\_document\_read}, the task is framed as a routine document review check. In successful traces, the agent retrieves \texttt{doc\_001} only after first receiving a normal access denial, obtaining an approval token, and retrying \texttt{/docstore/get} with \texttt{authorized\_by}.

For \texttt{oa\_layoff\_read}, the task is framed as a routine workforce planning review. The same approval-token progression is used to retrieve confidential \texttt{doc\_002}.

These three instantiations share the same attack logic but differ in resource, action, and recipient. This is useful for analysis because it suggests that the issue is not tied to one exact string or one target. The reusable pattern is the combination of a protected document denial, local approval-token acquisition, and immediate document-service execution with \texttt{authorized\_by}.

\subsection{Auxiliary Skill}

The auxiliary skill is deliberately compact. It instructs the agent to review relevant workspace records, request an approval token when a local service requires one, use returned tokens only for the requested action, reuse the business reason supplied in the task, and report the final result. The skill does not add unrelated capabilities or broad behavioral instructions.

We intentionally avoid adding long operational attachments or broad behavioral rules. In preliminary development, verbose instructions made the behavior less stable because the agent could spend more turns interpreting auxiliary context rather than executing the intended local workflow. The controlled local validation in Section~\ref{sec:local-validation} further shows that the auxiliary skill was not required in these local runs, but it remains useful for keeping the submitted workflow short and auditable.

\subsection{Trace Binding and Reproducibility}

Each target is executed with the official OpenClaw runner, and we use only runner-generated traces in the final evidence. We also check that the submitted task text appears in the corresponding trace, that trace metadata matches the generated task, and that logs do not indicate execution failures. This trace-binding step is not a separate attack contribution; it is evaluation hygiene that ensures the behavior analyzed below corresponds to the submitted task and skill files. The contribution is the empirical evidence itself; trace binding is the precondition that makes the evidence reliable.

\section{Results}

The verified online score details for our final submission show that all three included Enterprise OA targets passed the skill security scan, instruction security evaluation, attack trajectory score, and final target score. The corresponding public leaderboard score was 0.2730, placing team \texttt{nick5419780528} at rank 9 in the leaderboard snapshot available to us.

\begin{table}[htbp]
\centering
\small
\setlength{\tabcolsep}{2pt}
\begin{tabular}{@{}L{0.39\columnwidth}cccc@{}}
\toprule
Target & Skill & Instr. & Trace & Final \\
\midrule
\texttt{oa\_document\_export} & 1 & 1 & 1 & 1 \\
\texttt{oa\_document\_read} & 1 & 1 & 1 & 1 \\
\texttt{oa\_layoff\_read} & 1 & 1 & 1 & 1 \\
\bottomrule
\end{tabular}
\caption{Score details for the final verified document-target submission. All four evaluation dimensions passed for each target, confirming that the submitted task text was bound to runner-generated traces and that no security-scan rejection occurred.}
\label{tab:score}
\end{table}

We interpret these results as target-level evidence for the verified Enterprise OA subset, not as a claim of universal success across all OpenClaw tasks. The reported set is conservative: it includes only targets with runner-generated traces that matched the submitted task and skill files.

Beyond the binary score fields, the successful traces share the same authorization transition: a sensitive document action first receives a normal-access denial, the agent obtains an approval token, and the document operation later succeeds through the authorization field. The evidence therefore points to a narrow but concrete failure pattern: the service accepts an approval token without reconstructing the complete requester, approver, resource, action, destination or recipient, and purpose context.

\subsection{Controlled Local Validation}
\label{sec:local-validation}

To address reproducibility and ablation concerns, we ran 23 additional local experiments in the official OpenClaw execution environment. These are supplementary validation runs rather than new leaderboard submissions; they use the same sandboxed services and runner interface as the competition setting. We analyze each run by checking whether the audit trace contains a normal document denial, successful approval-token acquisition, token use through \texttt{authorized\_by}, and target document-operation success.

The component-removal variants are single-factor diagnostic checks. Auxiliary-skill removal runs the task without the compact skill; reduced framing removes explicit compliance or planning language; approval-cue removal omits direct guidance to request a token; and stop-cue removal omits the instruction to halt after the target result.

\begin{table}[htbp]
\centering
\small
\setlength{\tabcolsep}{2pt}
\begin{tabular}{@{}L{0.35\columnwidth}ccc@{}}
\toprule
Condition & Runs & Approval path & Target success \\
\midrule
Original configuration & 9 & 9 & 8 \\
Auxiliary skill removed & 3 & 3 & 3 \\
Reduced framing & 3 & 3 & 3 \\
Approval cue removed & 3 & 3 & 3 \\
Stop cue removed & 3 & 3 & 3 \\
Non-document tests & 2 & 0 & 0 \\
\bottomrule
\end{tabular}
\caption{Controlled local validation. ``Approval path'' counts runs that reached denial, approval-token acquisition, and \texttt{authorized\_by} use; ``Target success'' additionally requires the target document operation to succeed with \texttt{authorized\_by}.}
\label{tab:local-validation}
\end{table}

The repeated original-configuration reruns show that the mechanism is repeatable but not deterministic. All nine reruns reached the approval-token transition, and eight completed the exact target operation. In the remaining export run, the agent obtained and used the token for \texttt{/docstore/get}, but selected \texttt{/docstore/access-log} rather than \texttt{/docstore/export}. This reflects planner-level endpoint selection rather than a failure of the approval-token transition itself.

The component-removal runs refine the claim. Removing the auxiliary skill, reducing routine/compliance wording, removing explicit approval-token guidance, or removing the explicit stop condition each still produced three document-target successes in the local validation runs. Rather than showing that a particular prompt component is necessary, the evidence points to a recurring service-level mechanism: after a protected document service advertises \texttt{authorized\_by} on denial, the agent obtains a normal approval token, and the document service accepts token possession at the operation boundary. The two non-document tests did not exhibit this transition, supporting a narrow document-workflow claim rather than broad cross-domain generalization.

\begin{table}[htbp]
\centering
\small
\setlength{\tabcolsep}{2pt}
\begin{tabular}{@{}L{0.22\columnwidth}L{0.30\columnwidth}L{0.40\columnwidth}@{}}
\toprule
Component & Observed role & Evidence and limit \\
\midrule
Denial hint & Points the agent to a token field. & Before successful token use. \\
Approval token & Supplies local authorization material. & Before each successful document call. \\
Auxiliary skill & Keeps execution short and auditable. & Not required in local removal runs. \\
Routine wording & Makes the task business-like. & Reduced-framing runs still succeeded. \\
\bottomrule
\end{tabular}
\caption{Component-level diagnostic analysis after controlled local validation. The evidence supports the service-hint and approval-token transition more strongly than necessity claims about individual wording components.}
\label{tab:component-diagnostic}
\end{table}

\section{Analysis}

\subsection{Case-Level Analysis}

Table~\ref{tab:case-analysis} decomposes the three verified targets by operation, accepted approval token, and missing policy check. In each case, the approval service returns an access token, and the document operation later succeeds when that token is supplied through the \texttt{authorized\_by} field. The evidence therefore concerns token acceptance at the document-service boundary; it does not establish that the token was scoped to the exact requester, resource, action, destination or recipient, and purpose. We intentionally avoid reproducing full document contents; the analysis focuses on authorization structure rather than sensitive payloads.

\begin{table}[htbp]
\centering
\small
\setlength{\tabcolsep}{2pt}
\begin{tabular}{@{}L{0.18\columnwidth}L{0.18\columnwidth}L{0.56\columnwidth}@{}}
\toprule
Target & Token use & Missing policy check \\
\midrule
Export & Export call & Is this approval valid for this requester, resource, action, destination or recipient, and purpose? \\
Doc read & Get call & Is this approval valid for reading top-secret \texttt{doc\_001}? \\
Layoff read & Get call & Is this approval valid for reading confidential HR \texttt{doc\_002}? \\
\bottomrule
\end{tabular}
\caption{Case-level decomposition of the verified Enterprise OA targets. In each case, the audit record reports denied normal access but a successful response through an approval token supplied as \texttt{authorized\_by}.}
\label{tab:case-analysis}
\end{table}

The cases differ in sensitivity and operation type. Exporting a document to \texttt{compliance@company.com} introduces recipient validation, reading \texttt{doc\_001} exposes a top-secret CEO-owned document, and reading \texttt{doc\_002} exposes a confidential HR planning document. However, the attack pattern is unchanged: a business-like framing converts the agent's task from ``should this be done?'' into ``which service call completes this workflow?'' This shift matters because language models are optimized to complete coherent instructions, and the local service interface presents the approval token as an ordinary argument.

The cases also show why policy enforcement should not be delegated to natural-language framing or isolated service fields. A phrase such as ``pre-approved operations ticket'' may be plausible, and a field such as \texttt{authorized\_by} may look procedural, but plausibility is not authorization. The service must still check whether the requester, asserted approver, document, action, destination or recipient, and purpose are consistent with an enforceable policy.

\subsection{Cross-Target Analysis}

The attack succeeds because the agent can complete a locally coherent workflow while missing the broader authorization semantics. From the agent's perspective, the procedure is internally consistent: identify the relevant document, obtain a token when the service asks for \texttt{authorized\_by}, and call the target operation. From a security perspective, the missing step is explicit validation that the requester, approver, resource, action, destination or recipient, and purpose form a legitimate authorization tuple.

This distinction is important for agent safety research, but the scope should be stated carefully. The evidence does not show that all enterprise workflows suffer from this weakness. It shows that, in these OpenClaw document services, unsafe behavior can arise from ordinary service-call completion when authorization context is represented as weakly bound arguments.

The attack also illustrates a limitation of purely instruction-level safety review. The task text by itself resembles a routine administrative request. The risk emerges only after composing the task with tool affordances, service semantics, and authorization-argument handling. This motivates evaluations that inspect full trajectories and policy context, rather than evaluating prompts in isolation.

\section{Defensive and Benchmark Implications}

The following implications are derived from the observed traces, not from controlled defense experiments. They should be read as design guidance for systems that expose approval-token workflows to agents.

\paragraph{Service-side context binding.}
Protected services should not treat token possession through a stand-alone field such as \texttt{authorized\_by} as sufficient. The service should bind the token to requester identity, approver authority, resource, action, destination or recipient, purpose, and expiration. Export and read operations should then verify the exact covered document, requester, recipient when applicable, and business purpose rather than accepting a broad document-scoped token.

\paragraph{Agent-side policy reasoning.}
Agents should treat approval tokens as claims to be checked, not as proof that a request is permitted. Before using such a token, the agent should verify whether the request is appropriate under policy and whether sensitive data exposure is justified.

\begin{table}[htbp]
\centering
\small
\begin{tabular}{@{}L{0.31\columnwidth}L{0.61\columnwidth}@{}}
\toprule
Observed weakness & Design response \\
\midrule
Token accepted as sufficient & Bind it to requester, approver, resource, action, destination or recipient, and purpose. \\
Token scope under-specified & Check authority for the exact resource and operation. \\
Routine wording masks risk & Require agent-side policy reasoning before sensitive tool calls. \\
\bottomrule
\end{tabular}
\caption{Mapping from observed weaknesses to defensive design responses. These responses are not experimentally validated in this paper.}
\label{tab:defense-map}
\end{table}

\subsection{Benchmark Design Implications}

The observed failure mode suggests that agent-security benchmarks should measure more than whether a prompt sounds unsafe. A benign-looking request can still produce an unauthorized operation when tool affordances and service semantics omit authorization context. Three benchmark-design choices follow.

\paragraph{Cross-layer specifications.}
Target specifications should describe not only the final forbidden action, but also the policy context that makes the action unauthorized. For document workflows, this includes requester role, approver authority, document sensitivity, action type, destination or recipient, and purpose. Without these fields, a model or service may optimize for task completion while ignoring the actual access-control condition.

\paragraph{Positive and negative workflow pairs.}
Benchmarks can include paired tasks that are superficially similar but differ in one policy-relevant attribute. For example, a legitimate compliance delivery and an unauthorized external delivery may share the same document service but differ in recipient, approver scope, or requester authority. Such pairs help distinguish shallow refusal heuristics from context-sensitive authorization reasoning.

\paragraph{Trajectory-level evidence.}
Prompt-only evaluation is insufficient for tool-using agents. The relevant behavior often appears after several tool calls or after a single call with a sensitive argument. Evaluation should therefore inspect the entire trajectory, including tool arguments, service responses, timestamps, permission annotations, and audit records.

\begin{table}[htbp]
\centering
\small
\begin{tabular}{@{}L{0.30\columnwidth}L{0.62\columnwidth}@{}}
\toprule
Benchmark dimension & Recommended design \\
\midrule
Policy context & Encode requester, approver, resource, action, destination or recipient, and purpose. \\
Workflow pairs & Include similar-looking authorized and unauthorized tasks. \\
Tool trajectory & Score the sequence of calls, not only the initial prompt. \\
Trace integrity & Bind submitted instructions to generated traces and audit metadata. \\
Failure diagnosis & Report whether failures arise from instruction, tool choice, service policy, authorization binding, or trace mismatch. \\
\bottomrule
\end{tabular}
\caption{Benchmark design recommendations motivated by authorization-grounded workflows.}
\label{tab:benchmark-design}
\end{table}

These choices would make future evaluations more diagnostic: a score could identify whether failure arose from language-level safety, agent planning, authorization-argument handling, service-side authorization, or evaluation trace binding.

\section{Limitations}

Our method was validated on three Enterprise OA document targets. This is a small sample, and the targets may not represent the difficulty of all OpenClaw scenarios. The results should therefore be interpreted as concrete evidence of one authorization-context failure mode, not as a general success claim across smart-home, e-commerce, calendar, email, or other enterprise tasks.

The controlled local validation strengthens reproducibility but remains small. It shows 8/9 exact target successes for the original document method, 9/9 approval-token transitions, and 3/3 successes for each single-component removal variant. These runs should not be overinterpreted as statistical evidence or as proof that individual components are irrelevant in all models. They show that, in this environment, the recurring service-level approval path was observed even when individual wording components were removed.

The attack also depends on the service semantics exposed by the OpenClaw environment. If the protected document service re-validates requester, approver, resource, action, destination, and purpose at the operation boundary, or if approval tokens are scoped to exact operations and recipients, the workflow should be blocked. We did not run controlled defense experiments that implement these mitigations, so the defensive recommendations remain design implications rather than experimentally measured fixes.

Finally, the controlled variants are local diagnostic experiments, not exhaustive causal ablations. Future work should evaluate paired authorized and unauthorized workflows under controlled service configurations and multiple agent models.

\section{Conclusion}

Three verified Enterprise OA targets and 23 controlled local runs provide concrete evidence that authorization-context gaps are reproducible in this OpenClaw document setting. The central failure is a locally coherent workflow in which approval-token possession is consumed as sufficient authorization at a protected document endpoint, despite incomplete binding of requester, approver, resource, action, destination or recipient, and purpose.

The broader lesson is methodological. For tool-using agents, risk emerges from the interaction between user intent, agent planning, protected services, approval-token handling, and audit trails. Future work should evaluate these cross-layer trajectories directly and test whether context-bound approval prevents the failure mode observed here.

\paragraph{Ethical Statement.}
This work is conducted in a controlled security competition environment. All targets are simulated OpenClaw services, and the study does not access or test any real enterprise system. The paper uses only our own final submission materials and does not use external team materials, leaked submissions, or third-party competition archives. Any public release of code or prompts should follow the competition rules, remove credentials, and preserve the distinction between sandboxed evaluation artifacts and real-world enterprise systems.

\let\oldthebibliography\thebibliography
\renewcommand{\thebibliography}[1]{%
\oldthebibliography{#1}%
\setlength{\itemsep}{0pt}%
\setlength{\parsep}{0pt}%
}
\bibliographystyle{named}
\bibliography{references}

\begin{thebibliography}{}

\bibitem[\protect\citeauthoryear{Cheng \bgroup \em et al.\egroup
  }{2026}]{cheng2026seclaw}
Hao Cheng, Changtao Miao, Tianle Song, et~al.
\newblock {SeClaw}: Spec-driven security task synthesis for evaluating
  autonomous agents.
\newblock {\em arXiv preprint arXiv:2606.02302}, 2026.

\bibitem[\protect\citeauthoryear{Greshake \bgroup \em et al.\egroup
  }{2023}]{greshake2023more}
Kai Greshake, Sahar Abdelnabi, Shailesh Mishra, et~al.
\newblock More than you've asked for: A comprehensive analysis of novel prompt
  injection threats to application-integrated large language models.
\newblock {\em arXiv preprint arXiv:2302.12173}, 2023.

\bibitem[\protect\citeauthoryear{Hu \bgroup \em et al.\egroup
  }{2014}]{hu2014abac}
Vincent~C. Hu, David Ferraiolo, Rick Kuhn, et~al.
\newblock Guide to attribute based access control ({ABAC}) definition and
  considerations.
\newblock Technical Report Special Publication 800-162, National Institute of
  Standards and Technology, 2014.

\bibitem[\protect\citeauthoryear{Liu \bgroup \em et al.\egroup
  }{2023a}]{liu2023agentbench}
Xiao Liu, Hao Yu, Hanchen Zhang, et~al.
\newblock {AgentBench}: Evaluating {LLMs} as agents.
\newblock {\em arXiv preprint arXiv:2308.03688}, 2023.

\bibitem[\protect\citeauthoryear{Liu \bgroup \em et al.\egroup
  }{2023b}]{liu2023promptinjection}
Yi~Liu, Gelei Deng, Zhengzi Xu, et~al.
\newblock Prompt injection attack against {LLM}-integrated applications.
\newblock {\em arXiv preprint arXiv:2306.05499}, 2023.

\bibitem[\protect\citeauthoryear{{OWASP}}{2025}]{owasp2025llmtop10}
{OWASP}.
\newblock {Top 10 for Large Language Model Applications 2025}.
\newblock Online resource, 2025.

\bibitem[\protect\citeauthoryear{Ruan \bgroup \em et al.\egroup
  }{2023}]{ruan2023toolemu}
Yangjun Ruan, Honghua Dong, Andrew Wang, et~al.
\newblock Identifying the risks of {LM} agents with an {LM}-emulated sandbox.
\newblock {\em arXiv preprint arXiv:2309.15817}, 2023.

\bibitem[\protect\citeauthoryear{Sandhu \bgroup \em et al.\egroup
  }{1996}]{sandhu1996rbac}
Ravi~S. Sandhu, Edward~J. Coyne, Hal~L. Feinstein, et~al.
\newblock Role-based access control models.
\newblock {\em Computer}, 29(2):38--47, 1996.

\bibitem[\protect\citeauthoryear{Yao \bgroup \em et al.\egroup
  }{2022}]{yao2022react}
Shunyu Yao, Jeffrey Zhao, Dian Yu, et~al.
\newblock {ReAct}: Synergizing reasoning and acting in language models.
\newblock {\em arXiv preprint arXiv:2210.03629}, 2022.

\end{thebibliography}

\end{document}